\documentclass{article}
\usepackage[T1]{fontenc}
\usepackage[utf8]{inputenc}
\usepackage[lbd]{ismir}  % Remove the "submission" option for camera-ready version
\usepackage{amsmath,cite,url}
\usepackage{graphicx}
\usepackage{color}
\usepackage{booktabs}
\usepackage{placeins}
\usepackage{amssymb}% http://ctan.org/pkg/amssymb
\usepackage{pifont}% http://ctan.org/pkg/pifont
\usepackage{algorithm} % Added
\usepackage[noend]{algpseudocode} % Added
\usepackage{gensymb} % Added
\usepackage{siunitx} % Added
\usepackage{multirow} % Added
\usepackage{siunitx} % Added
\usepackage{kotex}
\usepackage{microtype} % Added
\usepackage{paralist} % Added
\usepackage{float} % added
\title{Designing a Multimodal Viewer for Piano Performance Analysis - a Pedagogy-First Approach}
\multauthor
  {Joonhyung Bae$^1$ \hspace{0.4cm} Hyeyoon Cho$^1$ \hspace{0.4cm} Kirak Kim$^1$ \hspace{0.4cm} Dawon Park$^2$}
  {{\bfseries Taegyun Kwon$^1$ \hspace{0.4cm} Yoon-Seok Choi$^2$ \hspace{0.4cm} Hyeon Hur$^2$ \hspace{0.4cm} Shigeru Kai$^3$}\\
  {\bfseries Yohei Wada$^3$ \hspace{0.4cm} Satoshi Obata$^3$ \hspace{0.4cm} Akira Maezawa$^3$ \hspace{0.4cm} Jaebum Park$^2$}\\
  {\bfseries Jonghwa Park$^2$ \hspace{0.4cm} Juhan Nam$^1$}\\
  $^1$ KAIST, Daejeon, South Korea\\
  $^2$ Seoul National University, Seoul, South Korea\\
  $^3$ YAMAHA, Hamamatsu, Japan\\
  {\tt\small \{jh.bae, hyeyooncho, kirak, ilcobo2, juhan.nam\}@kaist.ac.kr}\\
  {\tt\small \{dw\_park, dallas71, sunny000927, parkpe95, pianistpark\}@snu.ac.kr}\\
  {\tt\small \{shigeru.kai, yohei.wada, satoshi.obata, akira.maezawa\}@music.yamaha.com}
  }

\def\authorname{J. Bae, H. Cho, K. Kim, D. Park, T. Kwon, Y.-S. Choi, H. Hur, S. Kai, Y. Wada, S. Obata, A. Maezawa, J. Park, J. Park and J. Nam}

\begin{document}
\maketitle 

\begin{figure*}[t]
\centering
\includegraphics[width=\textwidth]{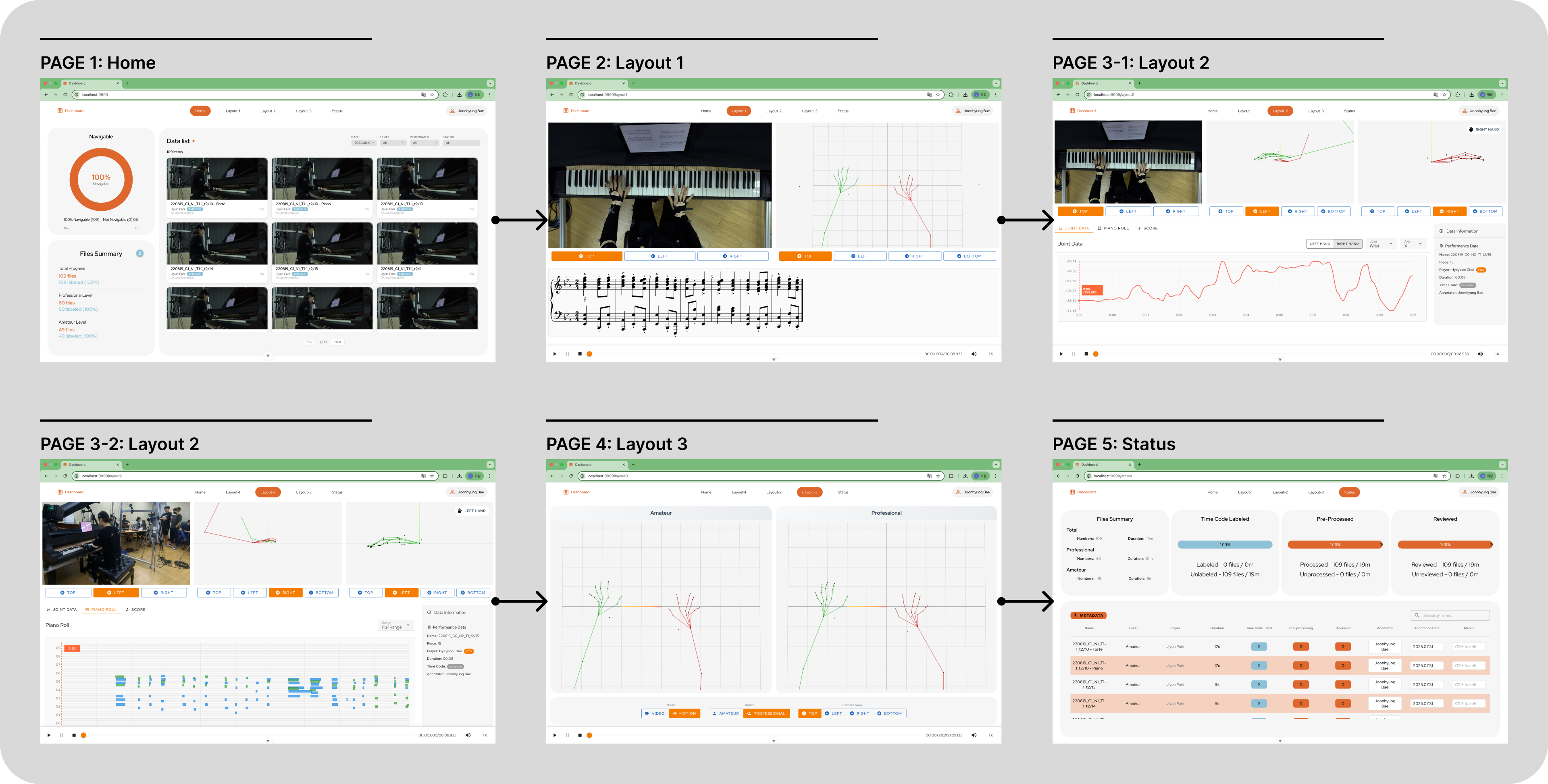}
\caption{Multimodal Piano Dataset Viewer System Overview - Integrated Analysis Environment with 5-Step Workflow}
\label{fig:system_overview}
\end{figure*}
\begin{abstract}

Abstract instructions in piano education, such as "raise your wrist" and "relax your tension," lead to varying interpretations among learners, preventing instructors from effectively conveying their intended pedagogical guidance. To address this problem, this study conducted systematic interviews with a piano professor with 18 years teaching experience, and two researchers derived seven core need groups through cross-validation. Based on these findings, we developed a web-based dashboard prototype integrating video, motion capture, and musical scores, enabling instructors to provide concrete, visual feedback instead of relying solely on abstract verbal instructions. Technical feasibility was validated through 109 performance datasets.
\end{abstract}

\section{Introduction and Related Work}

Abstract instructions in piano pedagogy, such as "relax your wrist," create a persistent communication barrier \cite{li2021teaching}. Existing technologies fail to address this barrier, as they either miss key physical gestures (Audio-MIDI analysis focuses on sound output rather than physical technique \cite{lerch2021interdisciplinary}), or focus too narrowly on note accuracy (commercial platforms primarily provide correctness feedback without addressing movement quality \cite{michalko2022toward}). While recent advances allow for rich multimodal data acquisition, including motion capture \cite{wang2024furelise}, a new bottleneck has emerged: a lack of clear frameworks for presenting this complex data in pedagogically meaningful ways \cite{nijs2010music}.

To address this presentation challenge, this paper proposes a user-centered design methodology. Instead of starting with technology, we derived core requirements from pedagogical practice, conducting in-depth interviews with a piano professor with 18 years of teaching experience. Based on these findings, we developed the Multimodal Piano Dataset Viewer, a web-based prototype specifically designed to meet these pedagogical needs.

This paper makes its core contribution by systematically defining seven expert-derived need groups for data presentation systems, demonstrating the technical feasibility of implementing these requirements in a web dashboard, and providing a concrete example of a pedagogy-first design methodology where educational needs, not technology, drive system development.

\section{From Pedagogy to Requirements}

We conducted an in-depth, semi-structured interview with a piano professor of 18 years, exploring limitations of current teaching methods and expectations for data presentation systems. Two researchers independently conducted thematic analysis \cite{braun2006using} on the transcribed interviews with cross-validation, deriving seven core need groups and associated design requirements. Based on this thematic analysis, we co-designed the system with a pianist, prioritizing features by feasibility and pedagogical impact to define the development scope.

\section{Core Design Requirements}
Through thematic analysis, we derived seven core need groups. \textbf{(1) In multimodal data integration}, the problem "it is difficult to analyze when viewing video separately, audio separately, and motion separately" was raised, requiring a synchronized integrated view of 3D motion + audio + scores. \textbf{(2) In intuitive communication}, the limitation "student comprehension varies greatly due to reliance on verbal explanations and analogies" was pointed out, necessitating concretization of abstract concepts through visualization. \textbf{(3) In detailed analysis tools}, the problem "it is difficult to reproduce performance once it has passed" required detailed analysis functions such as slow motion and section repetition. \textbf{(4) In technique and habit correction}, "it is difficult to correct wrong habits like hand angles and body usage," which led to expectations for objective feedback through 3D motion analysis. \textbf{(5) In efficient time management}, the constraint "it is difficult to view the entire piece when it is long" required quick navigation functions for efficient browsing of performance data. \textbf{(6) Objective evaluation} was expected to provide automatic analysis functions for identifying missed notes and ensuring fair judgment, but was set as scope for future research. \textbf{(7) Standardized terminology} raised the need for scientific systematization of abstract musical terms and was set as a long-term goal.

\section{Prototype Design and Implementation} 

We designed a five-stage workflow reflecting user needs, as illustrated in Figure~\ref{fig:system_overview}. \textbf{Page 1 - Home} allows confirmation of all data through video thumbnails, shows preprocessing progress, and presents explorable datasets. Filtering is possible by amateur/professional distinction, data collection date, and performer name, with access limited to data where Audio, MIDI, Video, and Motion are aligned. \textbf{Page 2 - Layout 1} places video on left, 3D motion visualization implemented with Three.js on right, and musical scores at bottom. Both video and motion allow switching between Top, Left, and Right viewpoints. Additionally, motion enables a Bottom viewpoint and free rotation, much like a 3D graphics program. \textbf{Page 3 - Layout 2} displays video, full-body motion, and motion separated for hand regions, with time-series line plots of joint-wise x,y,z coordinates, Piano Roll visualization, and musical scores at the bottom. Line Plot and Piano Roll display synchronized playback bars indicating the current playback position. \textbf{Page 4 - Layout 3} enables comparative analysis of amateur and professional performer hand motion or video on the left and right, with selective audio playback. We utilized multimodal data collected from one professional pianist and one amateur performer, with data collection methodology detailed in previous research \cite{kwon2023automated, park2023multivariate}. We confirmed that the five-stage workflow operates in a web environment and enables real-time synchronization of multimodal data, demonstrating the technical feasibility of core functions including synchronized view, section repetition, and quantitative analysis.

\section{Conclusion and Future Research}

This study addressed communication problems in piano education using an MIR-education convergence approach. Through systematic user research, we identified seven need groups and confirmed the technical feasibility of a web-based multimodal prototype. This demonstrates a user-needs-based design methodology over a technology-centered one, highlighting the educational potential of MIR technology. However, this study has limitations as preliminary research based on a single expert interview, which limits the generalizability of the findings. Being a prototype-level implementation, verification of actual educational effects was not performed. Important functions including overlay comparison between performers, automatic missed note detection, and DTW-based time synchronization were excluded from the first development phase. Future plans involve two main stages: first, expanding and validating user needs with more experts using the developed dashboard, and second, conducting user testing with instructors and learners to verify the system's educational effects. We expect this work will improve music education methodology.

\section{Acknowledgments}
This work was supported by Seoul National University Research Grant in 2021 and YAMAHA Corporation Research Grant.
\bibliography{ISMIR2025_LBD}
\end{document}